# PARTICLE SWARM OPTIMIZED POWER CONSUMPTION OF TRILATERATION

Hussein S. Al-Olimat[1], Robert C. Green II[2], Mansoor Alam[1], Vijay Devabhaktuni[1] and Wei Cheng[3]

[1]EECS Department, College of Engineering, University of Toledo, Toledo, OH, USA
[2]Department of Computer Science, Bowling Green State University, Bowling Green, OH, USA
[3]Department of Computer Science, School of Engineering, Virginia Commonwealth University, Richmond, VA, USA

*ABSTRACT*

*Trilateration-based localization (TBL) has become a corner stone of modern technology. This study formulates the concern on how wireless sensor networks can take advantage of the computational intelligent techniques using both single- and multi-objective particle swarm optimization (PSO) with an overall aim of concurrently minimizing the required time for localization, minimizing energy consumed during localization, and maximizing the number of nodes fully localized through the adjustment of wireless sensor transmission ranges while using TBL process. A parameter-study of the applied PSO variants is performed, leading to results that show algorithmic improvements of up to 32% in the evaluated objectives.*

## 1. INTRODUCTION

Wireless sensor networks (WSN) consist of many sensing devices which are distributed inside of a given area. Sensors in the network carry out different tasks such as recording weather conditions, sensing motion, or recording sounds in addition to many other tasks. In WSNs, sensors cooperate with each other to formulate a fully connected network to allow information sharing between the network nodes. Such networks have many applications both for civilian and military purposes, the position of sensing devices that record the humidity of a place or the position of a military vehicle in a war zone are two examples of such applications where knowing the location of the information source is very important.

Wireless sensor nodes in WSNs may be positioned permanently or dynamically in a field depending on the localization protocol and nodes functionalities as thoroughly discussed in [1]. For permanent localization scenarios, knowing the location of the sensor is not a problem throughout the life time of the network; but in dynamic networks, localizing nodes can be time and power consuming and, in some scenarios, a lack of accuracy may occur. To solve problems of localization accuracy and increase the number of localized nodes in a time critical localization scenarios, meta-heuristic solutions and novel range-based iterative localization algorithms have previously been proposed in [2–6]. Additionally, to allow mapping localization solutions into real world scenarios relaxations to the localization problem regarding the nodes ordering, anchor nodes distribution, or global information sharing were also discussed in [7, 8].

Trilateration-based localization (TBL) and Multilateration-based localization (MBL) techniques are among the well-known and most used methods for localization. In this study, the various performance aspects of the TBL algorithm are examined through the application of single and multi-objective variants of particle swarm optimization (PSO). We implemented three version of PSO in this study to allow nodes to vary the transmission power level when broadcasting messages during localization. Trade-offs between multiple objectives — the number of transmitted messages, number of localized nodes, power consumption and the time needed to localize as many nodes as possible — are studied.

However, for the sake of demonstrating the applicability of our methods, ranging and location estimations were both assumed as being correctly calculated with minimal errors, which means that this study do not really discuss the localization accuracy or signal noises. Instead, the methods of this study try to allow WSNs to reduce the overall power consumption of the localization process without affecting the localization time or localizability (i.e. the number of localized nodes). So the meta-heuristic methods implemented in this paper allow one to find optimal and balanced solutions in terms of energy consumption by minimizing the number of messages sent and localization time while trying not to negatively affect the localizability.

The paper present the results of three implemented versions of the particle swarm optimization and clearly show the performance of them while trying to optimize the WSN work. Additionally, it provides a complete parameter study that was formulated to enhance the performance of the multi-objective PSO (MOPSO).

The key novelty of this paper is the optimization of the power consumption of the whole network without the need to cluster or build any small sensor islands such as in [9] or in [5]. This study takes advantage of the functionalities of toady's WSNs nodes to enhance the performance of the whole network, the ZigBee technology of transceivers in wireless nodes made that possible by allowing us to use multiple transmission power levels, where the different variants of PSO were used to programmically change the transmission level after the evaluation of the designed fitness functions.

The next 4 sections contain the necessary background regarding WSNs, localization, power consumption, and PSO algorithm. Additionally, a complete parameter study and set of evaluations are also provided to examine the designed methods.

## 2. BACKGROUND

For this study, it is important to have an understanding of the basic form and function of WSNs, including potential issues regarding power consumption, as well as a clear understanding of the applied metaheuristic algorithm — PSO.

### 2.1. Power Consumption in WSNs

Wireless sensor nodes often use solar cells to extend the batteries life in order to allow nodes to run for longer times. Other methods of extending battery life are the intelligent slowing of power consumption through a reduction in listening time [10], increasing the sleep time [11], or modifying sampling rates [12]. Another method of accomplishing power reduction is the use of multiple transmission ranges as is seen in the well-known CC2420 ZigBee RF transceiver [13, 14]. CC2420 allow nodes to transmit messages using eight discrete output power levels, as discussed in section 28 of the transceiver data sheet [15].

Previous studies took the advantage of such functionalities and tried to optimize the power consumption by varying the output power as in [16], where the output power of nodes were varied based on the distance between the communicated nodes after sharing the information using RTS/CTS (Request to Send / Clear to Send) mechanism. Additionally, in [17] a localization protocol was proposed to optimize the power consumption after clustering nodes based on the used power levels. In most of the simulations and studies done in literature, they tried to change the power level sequentially while observing the effect of using different power levels, or assign power levels to nodes randomly. The two excellent previous references inspired us to use computational intelligence to vary the output power levels in order to find more stable and balanced solutions while actually being able to optimize the power consumption of WSNs.

Table 6.1 of the CC2420 data sheet list the five different modes in which the transceiver consumes different amount of power, the five modes are: Voltage regulator off (OFF); Power Down mode (PD); Idle mode (IDLE); Receive mode; and finally, the Transmit mode. In this study, however, we try to minimize the transmit power through minimizing the number of transmitted messages and minimize the average output power levels used by nodes. Depending on the localization protocol, all of the other modes including the transmit mode will be affected by the steps and behaviour of the localization procedure. For example, longer localization time means more power consumption due to the power consumed while in the idle mode. In order to make our methods as much protocol independent as we can and allow the methods to be used with any localization protocol, the methods will only optimized the Transmit mode of transceivers.

The Atmel AT86RF230 transceiver, as the CC2420, allow varying the output power not only between 8 levels, but between 16 discrete levels ranging from 3 dBm to 17.2 dBm [18]. Having more output power levels allow us to have minimum length of intervals than what the CC2420 can provide. In order to observe the effect of having multi output power levels, we designed three different PSO versions. The first two of them are binary PSO, single and multi-objective, where they were designed to vary the output power between three discrete levels only. On the other hand, the continuous multi-objective version was implemented to show the extreme case where we have infinite power levels; even though it is not possible by modern transceivers, in order to show how much optimization can actually be achieved having more power levels in hand. Next, we will discuss how we calculated the power consumption while nodes are in Transmit mode and using the discrete or the continuous output power levels.

### 2.1.1. Discrete Power Ranges

In this particular method transceivers were assumed to have three different power levels and wave length of each power level has been calculated. The allowable ranges were calculated based on work done by [14] as in (1) and (2), where $R$ is the range in meters, $P_o$ is the sender transmit power in dBm, $F_m$ is the fade margin in dB, $P_r$ is the receiver sensitivity in dBm, $f$ is the signal frequency in MHz, and $n$ is the pass-loss exponent.

$$R = 10^x \qquad (1)$$

$$x = (P_o - F_m - P_r + 30*n - 32.44 \\ -(10*n*Log[10]f))/(10*n) \qquad (2)$$

Three power levels were chosen to represent the minimum, medium, and maximum possible output power levels. The first power level transmit with -3 dBm ($\approx 0.5mW$) transmission power.

The second power level as the medium range power level with 1 dBm ($\approx 1.26 mW$), and the third power level as the maximum power level with 5 dBm ($\approx 3.16 mW$). The rest of the variables in the previous equations are taken from [14] where $F_m = 8$ dBm, $P_r = -98$ dBm, $f = 2405$ MHz and $n = 2.5$.

The three used power levels allowed transmission to a maximum range of: $132.22$; $91.47$; and $63.28$ meters, for the maximum, medium and minimum power levels, respectively. Transmission ranges are assumed to be perfect circles where the sender node is the center of the circle and the calculated distance is the radius of the circle. Also, knowing the transmission distance ($R$) allows the determination of which nodes can communicate directly through a 1-hop connection.

### 2.1.2. Continuous Transmission Ranges

In this method it is assumed that the maximum range a ZigBee transceiver can reach is 132 meters and the minimum is 60 meters. As in the discrete method before, the energy consumption is calculated based on [14], but in this method the transmission range varies continually, as in (3) and (4), rather than the power range before. Finally, (4) is used to convert the power consumption $P_o$ from $dBm$ to $mW$.

$$P_o = (10 * n * \log_{10} R) + (10 * n * \log_{10} f) \\ - (30 * n) + F_m + P_r + 32.44 \qquad (3)$$

$$\hat{P}_o = 10^{(\frac{P_o}{10})} \qquad (4)$$

### 2.2. WSNs and Localization

A typical WSN consists of $N$ sensor nodes scattered among a field of $M \times M$ meters. Each node has a transmission range of $R$ and may or may not be equipped with various sensors such as temperature or humidity sensors, or radios such as GPS. Each node also holds a state of being localized, i.e. aware of its own position in the global or local positioning system, or unlocalized, i.e. not aware of its own position in space. Each node in the WSN can eventually be localized with the help of three already localized neighbor nodes that a node can communicate with over 1-hop connections (thoroughly discussed and proved in [19]). Two nodes are said to have a 1-hop connection if the distance between them is less than or equal to the transmission range, $R$. The localization procedure is the step that precedes actual network transmissions which, in the long run, will help in data forwarding and routing procedures between nodes in the network [20].

RSSI is widely used in localization protocols for ranging and position estimation, therefore, a study by Chen and Terzis [21] was made to calibrate the raw RSSI values from transceivers in order to allow better distance estimations when using RSSI values of the received messages. The study took advantage of the multiple discrete output power levels of the transceivers to send messages between nodes which put the applicability of using multiple power levels and the RSSI under the test. On the other hand, according to a study by [10], the increase of the output power do not necessarily increase the distance a message can reach. But, a recent study conducted by [22] showed stability in the RSSI values and reliable range measurements while using external antennas on Z1 motes.

In this study, the TBL method is used to allow the unlocalized nodes (known as: blind nodes) to localize themselves based on the difference in distance between them and the already localized neighbor nodes [23]. As an example of this process, Fig. 1 shows two steps of a simple localization process. The process starts by flooding information from the already localized nodes,

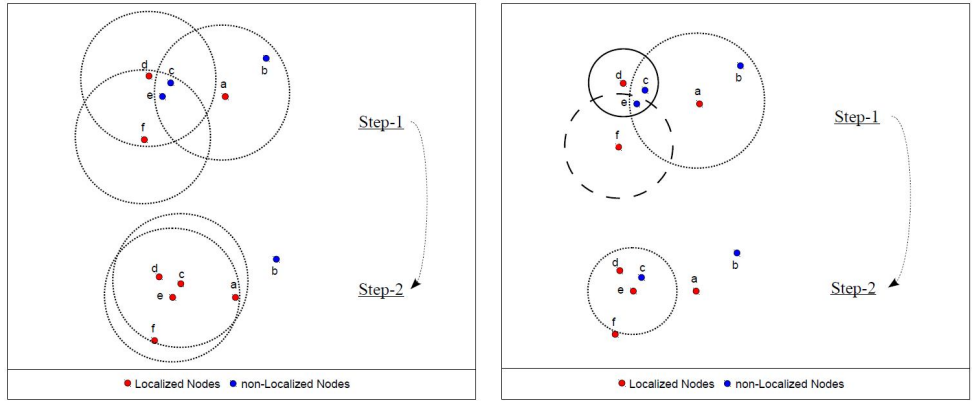

(a) Method 1: using one power level. The data of the steps are in Tables 1 and 2

(b) Method 2: using multiple power levels. The data of the steps are in Tables 3 and 4

Figure 1: A visual example of 2 steps in the TBL process

Table 1: TBL process using one power level (Step 1 Data)

| Nodes | a | d | f |
|---|---|---|---|
| Messages | 1 | | |
| Unit of Time | $1^{st}$ | | |
| Transmission Range | 132 | | |
| Power Consumption ($mW$) | 3.14 | | |
| **Non-Localized Nodes** | b | c | e |
| $\sum$ Received Messages | 1 | 3 | 3 |
| Localized? | No | Yes | |

Table 2: TBL process using one power level (Step 2 Data)

| Nodes | c | e |
|---|---|---|
| Messages | 1 | |
| Unit of Time | $2^{nd}$ | |
| Transmission Range | 132 | |
| Power Consumption ($mW$) | 3.14 | |
| **Non-Localized Nodes** | b | |
| $\sum$ Received Messages | 1 | |
| Localized? | No | |

in this case $a$, $d$ and $f$ where all nodes transmit using 132 meters transmission range which consumes 3.14 $mW$ per message sent, as calculated using (3) and (4). Step 1 of this localization process ends after flooding from the three nodes, where each of them broadcast one message that includes its global position within the $X - Y$ plane. The distance between all nodes can then be calculated from the RSS of the received messages. In this process and as a kind of relaxation to the problem, the events of broadcasting from senders and receiving by receivers were assumed to cost one unit of time, ignoring, for example, the possibility of collisions or retransmissions. Instead, each unit of time represent a step in the localization procedure where the units of time can later be mapped to actual localization scenarios using different protocols.

The Transmit mode power consumption of the three nodes after step-1 is: $3.14 \times 3 = 9.42 mW$, the complete details are listed in Table 1. By the end of Step-1, the broadcast messages are able to reach the three non-localized nodes, but only two of them was be able to localized themselves after receiving three messages. Nodes $c$ and $e$ are now considered localized after estimating their positions and, as all localized nodes, will contribute in step-2 of the localization process to help the other blind nodes. Meanwhile, node $b$ will keep waiting to receive two other messages to be able to localize itself.

In Step-2 of this localization process, nodes $c$ and $e$ broadcast, but none of their messages reach node $b$ which terminates the localization process since no new blind nodes were able to localize themselves. Nodes $a$, $d$ and $f$ are now finished participating in the localization process and will not broadcast again. The process will terminate after two time steps where, in this case, the nodes consumed 9.42 $mW$ in Step-1 and 6.28 $mW$ in Step-2 ($3.14 \times 2$ nodes = $6.28 mW$), leading to a cumulative power consumption of 15.7 $mW$.

It is clear that node $d$ is very close to both $c$ and $e$, which are the only blind nodes it can reach using the maximum transmission range, meaning that these nodes can minimize their power consumption by using a different, yet smaller, output power levels. The second method shown in Fig. 2 will allow each node to broadcast using a different power range, allowing each node to vary the reach distance of their messages. As listed in Tables 3 and 4, of the two steps of the localization, nodes $a, d, f$ and $e$ transmit using 132, 63.2, 91 and 83.4 meter transmission ranges respectively. Here, the transmission ranges were chosen randomly for the sake of demonstrating the multi-power level method, later, PSO will be the one choosing the different power levels leading to different transmission ranges.

Table 3: TBL process using multiple power levels (Step 1 Data)

| Nodes | a | d | f |
|---|---|---|---|
| Messages | 1 | | |
| Unit of Time | $1^{st}$ | | |
| Transmission Range | 132 | 63.2 | 91 |
| Power Consumption ($mW$) | 3.14 | 0.5 | 1.24 |
| **Non-Localized Nodes** | b | c | e |
| $\sum$ Received Messages | 1 | 2 | 3 |
| Localized? | No | | Yes |

Table 4: TBL process using multiple power levels (Step 2 Data)

| Nodes | e |
|---|---|
| Messages | 1 |
| Unit of Time | $2^{nd}$ |
| Transmission Range | 83.4 |
| Power Consumption ($mW$) | 1 |
| **Non-Localized Nodes** | b | c |
| $\sum$ Received Messages | 1 | 3 |
| Localized? | No | Yes |

The localization process, in this case, was not able to finish after two units of time as the previous example. Instead, it needs an extra one unit of time after to allow the new localized node in Step-2 ($c$) to broadcast. Assuming that node $c$ broadcasts using the maximum power range, the overall power consumption would be: Step-1 power consumption (3.14 + 0.5 + 1.24) $mW$; Step 2 power consumption (1 $mW$); and Step 3 power consumption (3.14 $mW$), which is equal to 9.02 $mW$. Thus, the main distinction between the outcomes of the two methods is the localization time and the power consumption. The second method was able to minimize the power consumption by 6.68 $mW$ while extending the localization time by adding an additional one unit of time or an extra step.

From the two examples it can be clearly seen that there is a trade-off between consuming more power while minimizing localization time and consuming less power but maximizing the localization time. This demonstrates the need to find better solutions by balancing the competing goals of maximizing the number of localized nodes (i.e. increase localizability), minimizing the overall power consumption (i.e. minimizing the output power levels and number of broadcast messages), and minimizing the localization time (i.e. minimizing localization steps). Consequently, the three PSO methods were designed to solve such problem and intelligently find better solutions.

### 2.3. Particle Swarm Optimization (PSO)

PSO is a population-based search algorithm inspired by bird flocking and fish schooling, where each particle learns from its neighbors and itself during the time it travels in space. The original single objective PSO (SOPSO) was first introduced by Kennedy and Eberhart in 1995 and operates over a continuous space [24]. Later, in 1997 a discrete, binary version of the algorithm was presented also by Kennedy and Eberhart to operate on discrete binary variables [25]. PSO was extended by Moore et al., as the first recorded attempt, to use PSO in a multi-objective problem (MOP) [26]. Later in 2002 multi-objective PSO (MOPSO) was introduced by [27] as an effective algorithm to solve MOPs.

All versions of PSO algorithm start by creating a number of particles to form a swarm that travels in the problem space searching for an optimum solution. An objective function should be defined to examine every solution found by each particle throughout the traveling time. A particle in this method is considered as a position in D-dimensional space, where each element can take a continuous value between a fixed upper and lower bounds. Additionally, each particle has a D-dimensional velocity, where each element also can take a bounded continuous value. Alternately, the elements of the positions matrix of the binary PSO can take the binary value of 0 or 1, where the value of each element of the velocity matrix is in the range [0, 1].

The individuals in PSO are a group of particles that move through a search space with a given velocity. At each iteration the velocity and position of each particle is stochastically updated by combining the particle's current solution, the particle's personal best solution or $\hat{p}_i$, and the global best solution or $\hat{g}$ over all particles (in MOPSO the procedure of choosing the best global is different, see the differences between SOPSO and MOPSO in the next pages). The required mathematics operating over a continuous space is listed in (5) to (9) where $\omega$ is the inertial constant, $c_1$ and $c_2$ represent cognitive and social constants that are usually $\sim 2$, and $r_1$ and $r_2$ are random numbers. $mRange$ and $xRange$ are the minimum and maximum transmission ranges respectively, and $Ran$ is a random number between 0 and $xRange$. (5) to (7) are used to update the velocity of the $i^{th}$ component of particle $p$, where (8) and (9) are used to update the position of that same component.

$$v_i = \omega v_i + c_1 r_1 \cdot (\hat{p}_i - p_i) + c_2 r_2 \cdot (\hat{g} - p_i) \quad (5)$$

$$\delta = \frac{(xRange - mRange)}{2} \quad (6)$$

$$v_i = \begin{cases} mRange & v_i < \delta \\ \delta & v_i \geq \delta \end{cases} \quad (7)$$

$$p_i = Ran + v_i \quad (8)$$

$$p_i = \begin{cases} mRange & p_i < mRange \\ xRange & p_i > xRange \\ p_i & mRange \leq p_i \leq xRange \end{cases} \quad (9)$$

Previous equations indicate that the velocity of neighbors and the current velocity of the particle itself will contribute in deciding the next position of the particle. In order for a particle to keep up with the other particles during the search of a solution, each particle adapts to the velocity of the swarm as a whole by learning from itself and its neighbor particles. Additionally, to improve the performance of the SOPSO, the inertia weight ($\omega$) in (5) can be modified dynamically (instead of a constant value) using mechanisms such as the simulated annealing to increase the probability of finding a near-optimal solution in fewer iterations and computing time.

Multi-objective Problems (MOPs) are known to have many contradictory objectives where enhancing the result of one objective will have a negative impact on the other objectives involved. MOPSO attempts to effectively find a solution or a set of solutions that ensure a balance between all the involved objectives as is thoroughly discussed in [28-36]. The main differences between the SOPSO and the MOPSO algorithms are:

1. MOPSO does not have a single global best solution, the $\vec{g}$ of the SOPSO in (5), that all particles learn from when they update their velocities in each iteration. Instead, MOPSO will have an archive of particles called leaders, where each leader is a potential solution of the problem. So instead of having only one global best solution the MOPSO will keep track of different solutions and use them randomly to lead other particles to update their velocities in each iteration using (5);
2. Dominance comparators are also implemented inside the MOPSO to help in finding a set of optimal solutions [37];
3. To avoid filling up the leaders archive, a crowding distance based on the non-dominated sorting genetic algorithm-II (NSGA-II) is used to decide which particles must remain in the archive [38, 39]; and
4. A mutation operator is applied to a portion of the swarm to improve the exploration and search ability and to avoid premature convergence [34, 35, 37]. Using the mutation method allowed us to give up using a simulated annealing method used to enhance the SOPSO performance by dynamically varying the inertia weight value.

## 3. PROPOSED APPROACH

The method proposed in this study involves the use of single- and multi-objective PSO to choose the appropriate, discrete or continuous output power levels for each wireless sensor node, in order to optimize various single or combinations of objectives including localization time, messages sent during localization, and power consumed. Appropriately, the remainder of this section will examine the use of the previously discussed three discrete power ranges and the continuous transmission ranges in the variants of the designed PSO.

### 3.1. PSO Formulation

In order to implement PSO, multiple objective functions as well as a problem specific representation are defined in the following sections. Three variants of PSO are presented including 1) A binary single objective PSO, 2) A binary multi-objective PSO, and 3) A continuous multi-objective PSO.

### 3.1.1. Objective Functions

As previously mentioned, the SOPSO and MOPSO algorithms are used in order to intelligently adjust the various discrete power ranges or the continuous transmission ranges of each sensor node. Accordingly, a representation consisting of *N* dimensions is used in order to represent each sensor that is deployed in the field. Furthermore, the objective functions for messages sent, time required for localization, power consumption, and number of nodes localized are calculated as follows:

International Journal in Foundations of Computer Science & Technology (IJFCST), Vol.4, No.4, July 2014

1. **Messages sent:** Depending on the localization procedure and communication mechanism between nodes, the number of messages sent back and forth between nodes will vary. However, in this study we assume that each already localized node will broadcast once in order to help other non-localized nodes achieve localization. Thus, the number of sent messages depends on the number of localized nodes.
2. **Time required:** In the proposed method, one unit of time is equivalent to one step in which sender nodes broadcast their locations and receivers receive the information. The step ends by running the location estimation using TBL method for each blind node that receives at least three messages from three different localized nodes. The localization procedure is going to terminate when no any new blind node was able to localize itself by the end of each step.
3. **Power consumption:** The variance in this objective mainly comes from the use of discrete and continuous transmission ranges, leading to various levels of power consumption. The power consumption is measured based on the power level or the transmission range each node uses to broadcast its message. Accordingly, the power consumption is the sum of each node's power consumption as chosen by PSO.

Table 5: Binary PSO Positions Matrix

| Range | Min | Mid | Max |
|---|---|---|---|
| $node_1$ | 0 | 1 | 0 |
| $node_2$ | 1 | 0 | 0 |
| $node_3$ | 0 | 0 | 1 |

Table 6: Continuous PSO Positions Matrix

|  | Transmission Range |
|---|---|
| $node_1$ | 83.4 |
| $node_2$ | 63.2 |
| $node_3$ | 91.0 |

4. **Number of nodes:** Choosing which power range a node will use to transmit messages or the transmission range of each node plays a significant role in the number of localized nodes. Through use of this objective, the proposed method maximizes the number of nodes capable of localizing using the least amount of consumed power, which means the least average transmission ranges of all nodes.

### 3.1.2. Binary PSO Representation

In all versions of PSO, each particle represents a potential solution of the localization problem. The variant of binary PSO (the single and multi-objective implementations) used in this study creates a random positions matrix where each element in the matrix takes the value of 0 or 1. Each row of the matrix represents a single node and consists of three columns corresponding to the min, mid, and max power ranges. An example of the matrix is shown in Table 5. In this particular example, nodes 1, 2, and 3 are assigned the mid, min, and max power ranges. The velocity matrix of each particle also look like the positions matrix but with continuous values ranges between 0 and 1.

### 3.1.3. Continuous PSO Representation

The variant of continuous MOPSO used in this study creates a random positions matrix where each element in the matrix takes a value between 60 to 132 meters (instead of a binary value) where each row of the matrix represents a single node's transmission range. An example of the matrix is shown in Table 6. In this particular example, nodes 1, 2, and 3 are assigned 83.4, 63.2 and 91.0 meters as the transmission ranges. The velocity matrix of each particle will also look like the positions matrix but with values between δ, calculated by (6), and the minimum transmission range value chosen ($mRange$), which is 60.

International Journal in Foundations of Computer Science & Technology (IJFCST), Vol.4, No.4, July 2014## 3.2. MOPSO

As discussed previously, the problem presented in this study is a multi-objective problem (MOP) and in order to accommodate these objectives, a PSO method capable of handling MOPs was implemented in order to find the trade-off between the contradicted objectives and to find a set of optimal solutions instead of finding only a single solution.

Algorithm 1 shows the pseudo-code of the two implemented MOPSO versions. The algorithm starts by initializing the swarm, in Line 2, where the positions and velocities matrices are initialized and the fitness values are calculated for all particles. In Line 3, the leaders archive is initialized from the swarm particles, but a series of checks must take place before adding a solution to the archive. A modified version of the implemented dominance and equality comparators of SMPSO [38] and OMOPSO [39] in JMetal [40] are used to make sure that no dominated solution is added to the archive. However, to calculate the crowding distances of leaders (Line 4), the exact JMetal method is used.

The swarm begins traveling in the problem space to find optimal solutions between Line 5 and Line 18. In Line 8, two leaders are chosen from the archive randomly and the crowding distance comparator compares the two leaders where the higher crowding distance will be chosen to assure diversity. Line 9 and 11 update the velocities and positions of each particle using (5) and (8), respectively.

In Line 12, a boundary mutation is implemented to mutate a portion of the swarm population. The flooding procedure and the localization process starts in Line 13 to measure the fitness of each particle's solution. In Line 14, the memory of the particle is updated to set the best fitness of the particle as the current fitness if it was considered as a better position by the comparators. Line 15 adds new particles to the leaders archive in the case that quality solutions are found. Finally, Line 16 does the same job as Line 4. All the initialization values of all parameters are described in the Parameter Study section.

```
Algorithm 1 MOPSO Algorithm
 1: procedure MOPSO(nodesList)
 2:     initialize swarm;
 3:     initialize leaders archive;
 4:     measure crowding distances;
 5:     for i ← 0 → numberIterations do
 6:         for j ← 0 → numberParticles do
 7:             procedure CALCULATENEWVELOCITIES
 8:                 choose random leader as global best;
 9:                 update velocities;
10:             end procedure
11:             calculate new positions;
12:             run MOPSO mutation;
13:             evaluate the solution;
14:             update particle memory;
15:             update leaders archive;
16:             measure crowding distances;
17:         end for
18:     end for
19: end procedure
```

## 4. SIMULATIONS AND RESULTS

### 4.1. Implementation

Fig. 2 shows the flow chart of the simulation procedure. Note that in Step-2, the implemented Java code reads the positions of each node from a saved topology file, where each node's types:

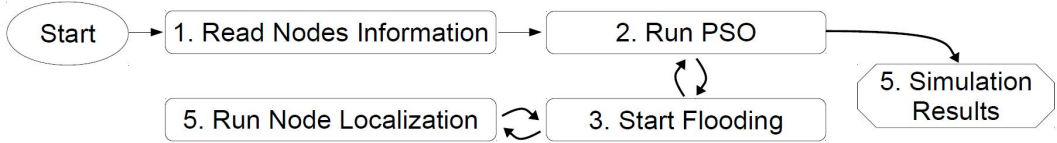

Figure 2: Simulation Flow Chart

anchor or blind node, in addition to the *X* and *Y* coordinates of the anchor nodes are saved. For this study a random WSN topology file containing 240 nodes, 40 of which are anchor nodes, are scattered among a field of 1000 × 1000 meters. In Step-3, one of the three proposed PSO versions is used. Step-4 and Step-5 is part of the fitness function where each particle's solution is examined by flooding the network and using the TBL localization method.

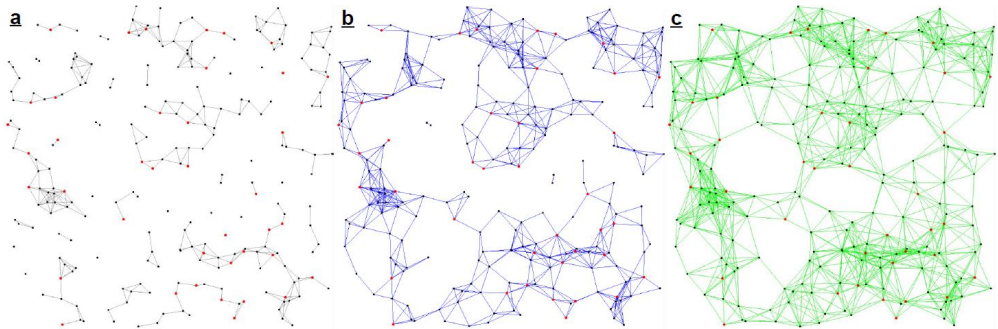

Figure 3: Simulation Panels showing 3 different simulation scenarios using the minimum (a), medium (b), and maximum (c) power levels. Lines between nodes represents a 1-hop connection between nodes and red nodes are anchor nodes.

### 4.2. Baseline Results

Initially, the WSN topology was examined three times using statically chosen power ranges. The networks created using these initial configurations are shown in Fig. 3 where anchor nodes are the red dots, normal nodes are the black dots, and the lines between nodes represents the 1-hop connection based on the chosen communication range.

As is listed in Table 7, the first run used only minimum power ranges for all the 240 nodes (Fig. 3-a) which allowed each node to transmit to a distance up to 63.28 meters. The localization procedure consumed 20.55 *mW*, which is the cost of transmitting 41 messages from the 40 anchors and the only one localized node by the anchors over 480 units of time. The second run used only medium range transmission (Fig. 3-b) which allowed each node to transmit to a distance up to 91.47 meters. After flooding, 96 nodes were localized with the help of the 40 anchor nodes and the localized nodes during 1,200 units of time through 136 messages which consumed 171.21 *mW*. In the last run (Fig. 3-c) all 200 nodes were localized where all nodes used the maximum power range during 960 units of time. The localization procedure consumed 758.95 *mW* of power when using the maximum power range where each node was able to transmit to a distance up to 132.22 meters.

From these baseline scenarios, it is clearly shown that using only one kind of power range may cause a large reduction in the number of localized nodes or consume excess energy to localize more nodes. The solution found by using the maximum allowed transmission is considered as the best solution in terms of the localization time and number of localized nodes. Yet, for power consumption, the solution using maximum transmission range is considered to be very poor. Note that in this baseline scenario and in the non-continuous versions of PSO the maximum allowed transmission range is 132.22 meters while in the continuous version the maximum is 132 meters.

Table 7: Baseline results using multi-power levels.

|  | $Run_1$ | $Run_2$ | $Run_3$ |
|---|---|---|---|
| Range | Min | Med | Max |
| Transmission Ranges | 63.28 | 91.47 | 132.22 |
| Time | 480 | 1,200 | 960 |
| Energy Consumption | 20.55 | 171.21 | 758.95 |
| Localized Nodes | 41 | 136 | 240 |

Table 8: SOPSO Parameters values

| Parameter | value |
|---|---|
| # Particles | 100 |
| # Iterations | 200 |
| Min Tran Range | 64 |
| Max Tran Range | 132 |
| $C_1$ and $C_2$ | 1.49445 |
| Inertia Weight ($\omega$) | RIW |

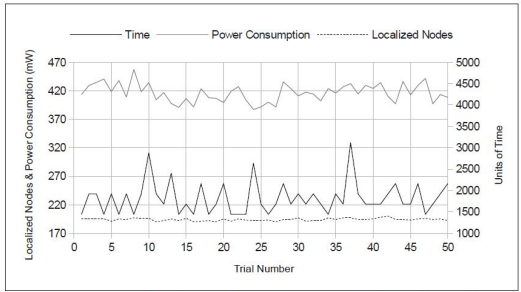
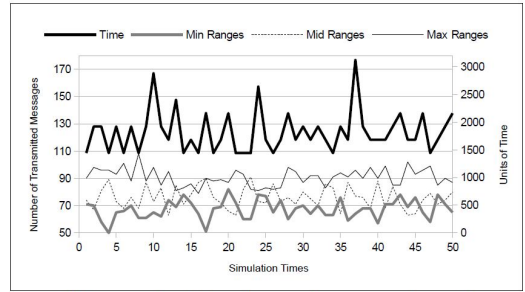

Figure 4: Results from 50 Trials of SOPSO while maximizing localizability.

Figure 5: Number of transmitted messages using different power levels.

**4.3. SOPSO**

Driven by the baseline solutions, SOPSO was used to explore the potential methods of improving the solutions to the presented problem. Accordingly, SOPSO was applied using each objective function previously discussed. The values of the SOPSO parameters are listed in Table 8, where a version of the random inertia weight proposed by [41] was implemented to make sure that PSO will not stuck in the local optima and improve the exploration ability by using a version of the simulated annealing method.

Fig. 4 shows 50 trials of simulating the WSN topology chosen, where the objective function was to maximize the number of the localized nodes only. In this scenario, the number of non-localized nodes ranged from 0 to 10 nodes, which means that the worst solution PSO was able to find was 5% worse in terms of the number of localized nodes. However, the power consumption was also reduced in addition to increasing the number of nodes localized. Yet, since the main focus of the trials was to maximize the number of localized nodes, the total time required for localization was worse, by 50% to 225%.

An obvious conclusion would be that whenever the larger power ranges are used, the overall power consumption will increase. Additionally, increasing the power ranges may increase the number of localized nodes. Fig. 5 suggests that a tremendous decrease in the number of lower power ranges used will aid in decreasing the localization time, representing a clear trade-off between these two objectives.

### 4.4. MOPSO Results

#### 4.4.1. Binary MOPSO

MOPSO was used to overcome the low quality of solutions found by SOPSO when minimizing the time needed for localization as well as power consumption while maximizing the number of localized nodes. The values of the MOPSO parameters are listed in Table 9, where a binary mutation method was implemented and a Fixed Inertia Weight (FIW) was chosen based on experiments discussed in the parameter study section 4.5.

The method was able to find a balance between all competing objectives and give solutions that outperforms the baseline and the SOPSO methods as detailed in Fig. 6, where the results of 50 trials are shown. It was able to find a balance between all of the competing objectives and, in some cases, it outperformed the two previous methods at all levels, (i.e. localizing all nodes during the shortest time possible and with power consumption less than any other solutions found by the methods before).

During the 50 trials, 115 different, yet optimal, solutions were found. Of these, 28 outperformed the baseline in terms of power consumption while maintaining the same time and number of localized nodes. The total power consumption ranged from 4% to 21% lower than the baseline measurements. In the best case, the MOPSO method improved power consumption by 29%, but was only capable of localizing 145 nodes — a clear trade-off.

Table 9: Binary and Continuous MOPSO Parameters values

| Parameter | Binary MOPSO | Continuous MOPSO |
|---|---|---|
| # Particles | 100 | 50 |
| # Iterations | 200 | 50 |
| Min Tran Range | 64 | 64 |
| Max Tran Range | 132 | 132 |
| Mutation Percentage | 15% | 20% |
| Mutation Value | Min Tran Range | Min Tran Range |
| $C_1$ and $C_2$ | 1.49445 | 1.49445 |
| Inertia Weight ($\omega$) | 0.1 | 0.1 |

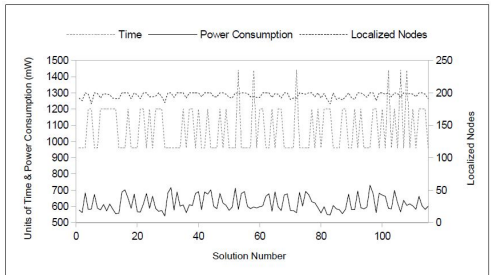
Figure 6: 115 solution from 50 trials of the MOPSO algorithm.

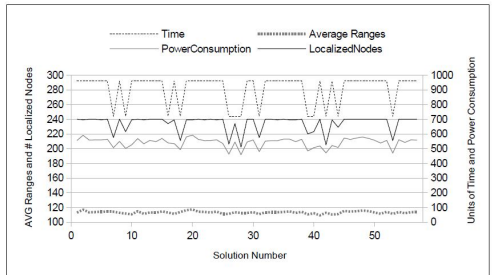
Figure 7: 57 solutions from 50 trials of the MOPSO algorithm.

#### 4.4.2. Continuous MOPSO

The main point of implementing this method was to enhance the quality of the solutions found by the binary MOPSO by using a continuous transmission range instead of 3 discrete levels. The values of the parameters used in the method are listed in Table 9, the values was chosen after a parameter study explained in Section 4.5. The method was able to find a balance between all competing objectives and give solutions that outperforms the other implementations as detailed in Table 10 and shown in Fig. 7.

During the 50 trials, 57 different, yet optimal, solutions were found. Of these, 53% outperformed the baseline in terms of power consumption while maintaining the same time and number of localized nodes (i.e localizing all nodes in 960 units of time). The total power consumption ranged from 24% to 32% lower than the baseline solutions. Fig. 8 shows how the continuous MOPSO outperforms the binary MOPSO method in terms of the average localization time which was decreased by up to 12.29%. Finally, the average number of localized nodes was worse by up to 0.72%.

An important observation to mention here is that the binary MOPSO tends to find more diverse solutions where some of solutions are of low quality, while the continuous version was able to find a set of more quality/balanced results, resulting in a lesser average number of localized nodes. The continuous MOPSO method tried to optimize the three other objectives by only making one of the objectives worse by as minimum as 0.72% which is reasonable in multi-objective problems.

The best solutions found by the two methods were when localizing all the nodes in less time and less power consumption. The best binary method solution consumed 600.97 *mW* to localize all the nodes in the shortest time, while the continuous method was able to minimize that by around 14% when the best solution was able to localized all the nodes with the same time by consuming only 517.73 *mW*. This clearly shows the advantage of using continuous transmission ranges instead of discrete ones.

Table 10: Averages and Standard Deviations of the continuous MOPSO

|  | Time | Power Consumption | Localized Nodes |
|---|---|---|---|
| AVG | 905.26 | 541.70 | 234.65 |
| STDEV | 100.70 | 32.03 | 10.52 |
| AVG+STDEV | 1005.96 | 573.73 | 245.17 |
| AVG-STDEV | 804.56 | 509.67 | 224.13 |

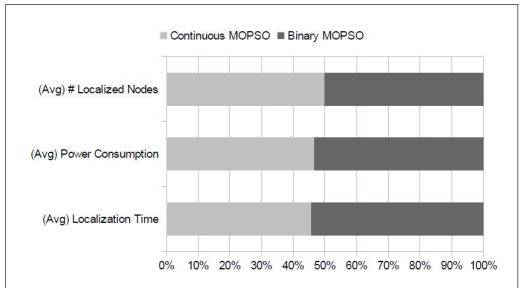

Figure 8: Results of the Binary vs. Continuous MOPSO

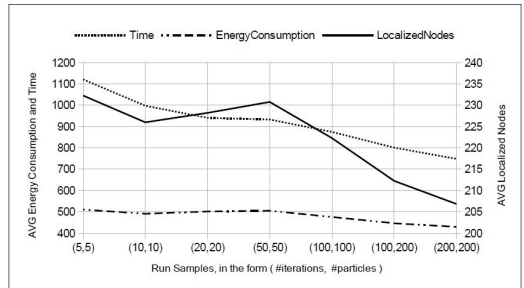

Figure 9: Experiments Varying The Number of Particles and Iterations

## 4.5. Parameter Study

Parameter studies, in general, are concerned with tuning the values of the elements that are involved in finding a solution. PSO is no exception and as was mentioned previously, PSO has many constant parameters that affect the global and local navigation abilities which in return affect the speed and direction of the particle. This study is based on the implemented continuous MOPSO method.

To achieve faster convergence and performance stability when executing the algorithm, a series of experiments was run while varying the value of 6 key parameters. Each time an experiment was run, the value of only one of the parameters was changed while not changing the values of the others. Additionally, each time a value changed, the algorithm was executed 50 times and the average results were calculated. The values of each experiment were then collected to measure the behavior of the algorithm where we generally measured the quality of the solution by comparing the average values of the 4 main objectives across all experiments. The results show a clear trade-off between the competing objectives so the values of each parameter was carefully chosen based on the quality and stability of solutions. Below is a list of the parameters with the results of each experiment:

1. **Number of Particles and PSO iterations:** As shown in Fig. 9, seven different combinations of the number of particles and iterations were used with the values varied from 5 to 200. Normally, we expect more improvement when increasing the value of the two parameters, and worse solutions when decreasing them. After comparing the average values of the solutions, it was found that using a swarm larger than 100 particles and running the PSO for more than 100 iterations will not improve the quality of the solutions. Additionally, using smaller swarms with lesser number of iterations negatively affected the quality of the solutions.

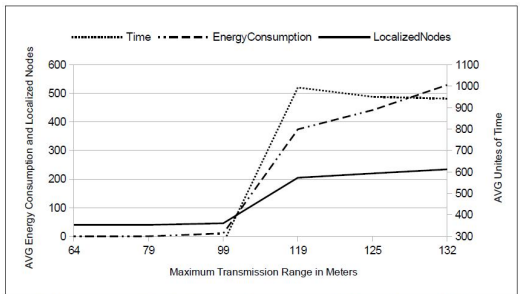
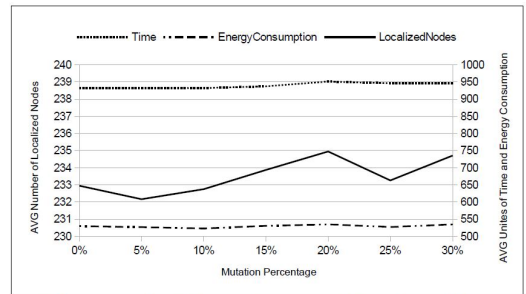

Figure 10: Experiments Varying The Maximum Transmission Ranges

Figure 11: Experiments Varying The Mutation Percentage

The values of the two parameters were set to 50 iterations and 50 particles due to the fact that when using this combination, a better stability than the other combinations was achieved, with standard deviation measured as up to 60% better than the (5,5) experiment, for example. Also, localizability was given the priority when measuring the quality of the solution, so the greatest average of localized nodes was chosen, but without affecting the other objectives. So, for example, (50, 50) was chosen over (20, 20) for the fact that in (50, 50) more nodes were localized in the same units, but with slightly more power — a reasonable result.

2. **Minimum and Maximum transmission ranges:** Transmission ranges are the most important parameter in this study as the focus was to minimize the average transmission ranges used by all nodes. ZigBee was designed to transmit data on no lower than -3 dBm which means the shortest transmission range of ZigBee is 62.68 meters and, thus, 64 was chosen as the lowest transmission power. For the maximum transmission ranges, it was found that using greater transmission ranges was able to maximize localizability, and if all nodes are localizable the algorithm is able to localize all of them a majority of the time. On the other hand, using smaller transmission ranges consume less power but gives poor results. Where In some cases the algorithm was not able to localize any node, as shown in Fig. 10.

Using large or small transmission ranges play a significant role in finding quality results. Generally, PSO will pick the values of the transmission ranges of each node from a continuous interval where the upper and lower bounds are the maximum and minimum ranges. We varied the upper bound from 64 to 132 meters while maintaining the same lower bound as 64 meters. Using maximum power ranges as 119-132 increased the average number of localized nodes from 412% to 418% more, i.e from 0 to as maximum as 233.9 of the 240 nodes. Obviously the time and energy consumption increased due to the fact we were able to localized more nodes. For range, values of 132 and 125 meters were found to be the best values, but 132 was chosen as the maximum transmission range because priority was given to the number of localized nodes which slightly increased the power consumption but was able to decrease the localization time and increase the number of localized nodes.

3. **Mutation operator:** The boundary mutation is used to avoid premature convergence by improving the search ability of the swarm. This method was adjusted by changing the mutation percentage in addition to the value of mutation. The mutation percentage was first varied from 0 to 30% during 7 experiments as shown in Fig. 11, then the mutation value was swapped between the chosen minimum and maximum transmission as shown in Fig. 12.

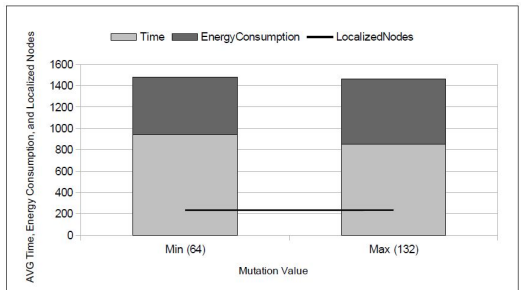
Figure 12: Experiments Varying The Mutation Value

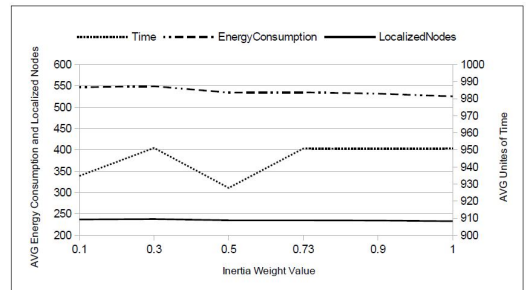
Figure 13: Experiments Varying The Inertia Weight

From Fig. 11, it can clearly be seen that 20 and 30% are the best choice in terms of the average number of localized nodes. 20% was chosen over the 30% mark for the fact that the 20% mutation percentage was found to be more stable with a standard deviation ranging from 12% to 22% less than the 30% mutation. The 20% mutation gave slightly better solutions (by around 0.10%), but the standard deviation of the 30% was found to be around 6% less than the 20% mutation with mutation value of 8.82, while the standard deviation for the 20% mutation was recorded as 9.38.

Fig. 12 shows how when using minimum transmission range as the mutation percentage the power consumption was found to be less without really affecting the number of localized nodes. However, reducing the power consumption when using the minimum transmission range increased the localization time but reduced the power consumption around 13%. We chose the minimum transmission range as the mutation value to prevent the MOPSO of using larger transmission ranges in order to minimize the power consumption as much as we can.

4. **PSO local weight ($c_1$) and global weight ($c_2$):** The two parameters are used in (5) in updating the velocity matrix to optimize the behavior of particles. Based on a study by Eberhart and Shi [42] that compared inertia weights and constriction factors in particle swarm optimization, the values of the acceleration coefficients, $c_1$ and $c_2$ in (5), were set to 1.49445. This was found to have better influence on the performance of PSO and was thus used.

5. **Inertia weight ($\omega$):** It is one of the most important adjustable parameters in PSO. $\omega$ value can impact the overall performance of the algorithm in finding a potential optimal solution in less computing time. In the proposed method a fixed inertia weight (FIW) value of $\omega$ is used as stated in (5). It is a fixed constant that is defined before running the algorithm. The value of it was varied from 0.1 to 1 as shown in Fig. 15 and used in many studies [41-45]. $\omega$ was set to 0.1, instead of the other relatively good choice 0.5. The value of 0.1 was chosen for the fact that it showed more stable results (up to 21.5% less standard deviation than the 0.5 level). Additionally, a value of 0.1 localized 0.7% more nodes than the 0.5, but increased the power and time by 2.3% and 0.75% respectively.

## 6. CONCLUSIONS AND FUTURE WORK

This paper has presented single-objective and multi-objective PSO-based solutions for the power consumption of WSN during trilateration-based localization. The overall performance of the TBL algorithm was evaluated and improved through the simultaneous optimization of various objective functions. Results clearly show that the use of SOPSO and MOPSO to optimize the TBL algorithm in terms of power consumption is effective, providing improvements up to 32% only on the Transmit mode of transceivers. Also, as shown by the study, using single global output power is less stable in localizing nodes than using multiple levels and using the maximum possible output level is not cost effective solution to the stability of localization, therefore, PSO was found to solve the problem without negatively affect the TBL work in terms of localizability in particular. However, our study can be mapped to real test beds using techniques such as the component based localization, nodes clustering, and RTS/CTS methods, in addition to many others, as suggest by [7, 8, 17].

The work conducted in this study will be used in future implementations of different localization protocols using Contiki-OS and real world motes such as the Z1 mote. The results of such more realistic simulations and implementations will allow us to verify the performance and quality of solutions achieved by this study.

## ACKNOWLEDGEMENTS

This material is based upon work supported in part by the National Science Foundation under Grant No. CNS-1248381.

## REFERENCES

[1] I. Amundson and X. D. Koutsoukos, \A survey on localization for mobile wireless sensor networks," in Proceedings of the 2Nd International Conference on Mobile Entity Localization and Tracking in GPS-less Environments, ser. MELT'09. Berlin, Heidelberg: Springer-Verlag, 2009, pp. 235-254. [Online]. Available: http://goo.gl/Cx9F4

[2] S. Zhang, G. Li, W. Wei, and B. Yang, \A Novel Iterative Multilateral Localization Algorithm for Wireless Sensor Networks," Journal of Networks, vol. 5, no. 1, pp. 112-119, Jan. 2010. [Online]. Available: http://goo.gl/4SHQr

[3] R. Kulkarni, \Bio-inspired node localization in wireless sensor networks," IEEE International Conference on Systems, Man, and Cybernetics, no. October, pp. 205-210, 2009. [Online]. Available: http://goo.gl/iIoxv

[4] A. Kumar, A. Khosla, J. S. Saini, and S. Singh, \Meta-heuristic range based node localization algorithm for Wireless Sensor Networks," 2012 International Conference on Localization and GNSS, pp. 1-7, Jun. 2012. [Online]. Available: http://goo.gl/7y3cr

[5] W. Cheng, N. Zhang, and M. Song, \Time-bounded essential localization for wireless sensor networks," IEEE Transactions on Networking, pp. 1-14, 2010. [Online]. Available: http://goo.gl/P8BnS

[6] D. K. Goldenberg, \Fine-grained localization in sensor and ad-hoc networks," Ph.D. dissertation, New Haven, CT, USA, 2006, aAI3243632.

[7]  D. K. Goldenberg, P. Bihler, Y. R. Yang, M. Cao, J. Fang, a. S. Morse, and B. D. O. Anderson, \Localization in sparse networks using sweeps," Proceedings of the 12th annual international conference on Mobile computing and networking – MobiCom'06, p. 110, 2006.

[8]  X. Wang, J. Luo, Y. Liu, S. Li, and D. Dong, \Component-Based Localization in Sparse Wireless Networks," IEEE/ACM Transactions on Networking, vol. 19, no. 2, pp. 540-548, Apr. 2011.

[9]  H. Ali, W. Shahzad, and F. A. Khan, Wireless Sensor Networks and Energy E_ciency, N. Zaman, K. Ragab, and A. B. Abdullah, Eds. IGI Global, Jan. 2012. [Online]. Available: http://goo.gl/ERT98k

[10] M. Mallinson, P. Drane, and S. Hussain, \Discrete Radio Power Level Consumption Model in Wireless Sensor Networks," 2007 IEEE Internatonal Conference on Mobile Adhoc and Sensor Systems, pp. 1-6, Oct. 2007. [Online]. Available: http://goo.gl/YU8yf

[11] J. Robles, S. Tromer, M. Quiroga, and R. Lehnert, \A low-power scheme for localization in wireless sensor networks," International Federation for Information Processing, pp. 259-262, 2010. [Online]. Available: http://goo.gl/H7FRd

[12] M. Bhuiyan, G. Wang, J. Cao, and J. Wu, \Energy and Bandwidth-E_cient Wireless Sensor Networks for Monitoring High-Frequency Events," Proc. of IEEE SECON, 2013. [Online]. Available: http://goo.gl/i8Rws

[13] W. C. Everywhere, \Wirelessly Connecting Everywhere," Wireless Connectivity, no. 2Q, pp. 1-72, 2013. [Online]. Available: http://goo.gl/tmD3o

[14] S. Farahani, ZigBee wireless networks and transceivers. Elsevier, 2011. [Online]. Available: http://goo.gl/3pWmQ

[15] T. I. Inc, \CC2420 2.4 GHz IEEE 802.15.4 / ZigBee-ready RF Transceiver Applications," Texas Instruments Inc, Dallas, Texas, USA., Tech. Rep., 2014.

[16] M.-H. Meng, \Power Adaptive Localization Algorithm for Wireless Sensor Networks Using Particle Filter," IEEE Transactions on Vehicular Technology, vol. 58, no. 5, pp. 2498-2508, 2009. [Online]. Available: http://goo.gl/khuNA

[17] K.-B. Chang, Y.-B. Kong, and G.-T. Park, \Clustering algorithm in wireless sensor networks using transmit power control and soft computing," in Intelligent Control and Automation, ser. Lecture Notes in Control and Information Sciences, D.-S. Huang, K. Li, and G. Irwin, Eds. Springer Berlin Heidelberg, 2006, vol. 344, pp. 171-175.

[18] Atmel Corporation, \AVR2001: AT86RF230 Software Programmer's Guide," Tech. Rep., 2007.

[19] J. Aspnes, T. Eren, D. K. Goldenberg, A. S. Morse, W. Whiteley, Y. R. Yang, B. D. O. Anderson, and P. N. Belhumeur, \A theory of network localization," IEEE Transactions on Mobile Computing, vol. 5, no. 12, pp. 1663-1678, Dec. 2006.

[20] Z. Hu, D. Gu, Z. Song, and H. Li, \Localization in wireless sensor networks using a mobile anchor node," in Advanced Intelligent Mechatronics, 2008. AIM 2008. IEEE/ASME International Conference on, July 2008, pp. 602-607.

[21] W.-n. Chen and J. Zhang, \An Ant Colony Optimization Approach to a Grid Workow Scheduling Problem With Various QoS Requirements," IEEE TRANSACTIONS ON SYSTEMS, MAN, AND CYBERNATICS, vol. 39, no. 1, pp. 29-43, 2009. [Online]. Available: http://goo.gl/f0P8xR

[22] M.-P. Uwase, N. Long, J. Tiberghien, K. Steenhaut, and J.-M. Dricot, \Poster abstract: Outdoors range measurements with zolertia z1 motes and contiki," in Real-World Wireless Sensor Networks, ser. Lecture Notes in Electrical Engineering, K. Langendoen, W. Hu, F. Ferrari, M. Zimmerling, and L. Mottola, Eds. Springer International Publishing, 2014, vol. 281, pp. 79-83. [Online]. Available: http://goo.gl/XaVJzm

[23] Francisco Sant, \Localization in wireless sensor networks," ACM Journal, vol. V, no. November, pp. 1{19, 2008.

[24] J. Kennedy and R. Eberhart, \Particle swarm optimization," Proceedings of ICNN'95 - International Conference on Neural Networks, vol. 4, pp. 1942-1948,1995. [Online]. Available: http://goo.gl/Srnox

[25] J. Kennedy and R. C. Eberhart, \A DISCRETE BINARY VERSION OF THE PARTICLE SWARM ALGORITHM," IEEE international conference on Systems, Man, and Cybernetics, pp. 4104-4108, 1997.

[26] J. Moore and R. Chapman, \Application of particle swarm to multiobjective optimization," Department of Computer Science and Software Engineering Department, Auburn University, pp. 1-4, 1999. [Online]. Available: http://goo.gl/NPkun

[27] K. E. Parsopoulos and M. N. Vrahatis, \Particle swarm optimization method in multiobjective problems," Proceedings of the 2002 ACM symposium on Applied computing - SAC '02, vol. 1, p. 603, 2002. [Online]. Available: http://goo.gl/XqAUZ

[28] C. a. Coello Coello and M. Reyes-Sierra, \Multi-Objective Particle Swarm Optimizers: A Survey of the State-of-the-Art," International Journal of Computational Intelligence Research, vol. 2, no. 3, pp. 287-308, 2006. [Online]. Available: http://goo.gl/Y0G0u

[29] Y. Wang and Y. Yang, \Particle swarm optimization with preference order ranking for multi-objective optimization," Information Sciences, vol. 179, no. 12, pp. 1944-1959, May 2009. [Online]. Available: http://goo.gl/nj34N

[30] H. S. Urade and R. Patel, \Dynamic Particle Swarm Optimization to Solve Multi-objective Optimization Problem," Procedia Technology, vol. 6, pp. 283-290, Jan. 2012. [Online]. Available: http://goo.gl/u2OOY

[31] S.-J. Tsai, T.-Y. Sun, C.-C. Liu, S.-T. Hsieh, W.-C. Wu, and S.-Y. Chiu, \An improved multi-objective particle swarm optimizer for multi-objective problems," Expert Systems with Applications, vol. 37, no. 8, pp. 5872-5886, Aug. 2010.

[32] D. Y. Sha and H. Hung Lin, \A particle swarm optimization for multi-objective shop scheduling," The International Journal of Advanced Manufacturing Technology, vol. 45, no. 7-8, pp. 749-758, Feb. 2009. [Online]. Available: http://goo.gl/OTAZf

[33] B. Alatas and E. Akin, \Multi-objective rule mining using a chaotic particle swarm optimization algorithm," Knowledge-Based Systems, vol. 22, no. 6, pp. 455-460, Aug. 2009. [Online]. Available: http://goo.gl/WU8P0

[34] S. Pang, H. Zou, W. Yang, and Z. Wang, \An Adaptive Mutated Multi-objective Particle Swarm Optimization with an Entropy-based Density Assessment Scheme," Information & Computational Science, vol. 4, pp. 1065-1074, 2013. [Online]. Available: http://goo.gl/IVsV1

[35] L. Wang, W. Ye, X. Fu, and M. Menhas, \A modi_ed multi-objective binary particle swarm optimization algorithm," Advances in Swarm Intelligence, pp. 41-48, 2011. [Online]. Available: http://goo.gl/wyHyt

[36] R. Eberhart, \Multiobjective optimization using dynamic neighbourhood particle swarm optimization," Proceedings of the 2002 Congress on Evolutionary Computation. CEC'02 (Cat. No.02TH8600), vol. 2, pp. 1677-1681, 2002. [Online]. Available: http://goo.gl/Iuh8L

[37] C. A. Coello, G. T. Pulido, and M. S. Lechuga, \Handling multiple objectives with particle swarm optimization," Trans. Evol. Comp, vol. 8, no. 3, pp. 256-279, Jun. 2004. [Online]. Available: http://goo.gl/Hhd2J

[38] a.J. Nebro, J. Durillo, J. Garcia-Nieto, C. Coello Coello, F. Luna, and E. Alba, \SMPSO: A new PSO-based metaheuristic for multi-objective optimization," 2009 IEEE Symposium on Computational Intelligence in Milti-Criteria Decision-Making, no. 2, pp. 66-73, Mar. 2009. [Online]. Available: http://goo.gl/LbGty

[39] M. Sierra and C. Coello, \Improving PSO-Based multi-objective optimization using crowding, mutation and -dominance," Evolutionary Multi-Criterion Optimization, 2005. [Online]. Available: http://goo.gl/cEaYBE

[40] J. J. Durillo and A. J. Nebro, \jMetal: A Java framework for multi-objective optimization," Advances in Engineering Software, vol. 42, no. 10, pp. 760-771, Oct. 2011. [Online]. Available: http://goo.gl/tZ5Rw

[41] L. Chong-min, G. Yue-lin, and D. Yu-hong, \A New Particle Swarm Optimization Algorithm with Random Inertia Weight and Evolution Strategy," Journal of Communication and Computer, vol. 5, no. 11, pp. 42-48, 2008.

[42] R. Eberhart and Y. Shi, \Comparing inertia weights and constriction factors in particle swarm optimization," Proceedings of the 2000 Congress on Evolutionary Computation. CEC00 (Cat. No.00TH8512), vol. 1, no. 7, pp. 84-88, 2000. [Online]. Available: http://goo.gl/YJbz23

[43] S. Sivanandam and P. Visalakshi, \Multiprocessor scheduling using hybrid particle swarm optimization with dynamically varying inertia," International Journal of Computer Science & Applications, vol. 4, no. 3, pp. 95-106, 2007. [Online]. Available: http://goo.gl/52UXL

[44] K. Deep and Madhuri, \Application of globally adaptive inertia weight pso to lennard jones problem," in Proceedings of the International Conference on Soft Computing for Problem Solving (SocProS 2011) December 20-22, 2011, ser. Advances in Intelligent and Soft Computing, Springer India, 2012, vol. 130, pp. 31-38.

[45] R. Ojha and M. Das, \An Adaptive Approach for Modifying Inertia Weight using Particle Swarm Optimisation," IJCSI International Journal of Computer Science Issues, vol. 9, no. 5, pp. 105-112, 2012. [Online]. Available: http://goo.gl/IAEnB.